\documentclass[5p,times,twocolumn]{elsarticle}
\usepackage{times,amsmath,amssymb,amsfonts,graphicx,bbm}
\usepackage{lineno}
\usepackage{color}
\journal{Physics Letters A}
\newcommand{\ket}[1]{\vert #1 \rangle} %
\newcommand{\braket}[2]{\langle #1 \vert #2 \rangle}
\newcommand{\ketbra}[2]{\vert #1 \rangle \! \langle #2 \vert}
\newcommand{\expec}[2]{ \left[ #1 \right]_{#2}}
\newcommand{\nnum}{\nonumber \\}
\def\ml{{\scriptscriptstyle M\!L}}
\def\opt{{\scriptscriptstyle M}}
\def\mx{{\scriptscriptstyle M}}

\begin{document}
\begin{frontmatter}
\title{Characterization of classical Gaussian processes using quantum probes}
\author{Claudia Benedetti, Matteo G. A. Paris}
\address{Dipartimento di Fisica, Universit\`a degli 
Studi di Milano, I-20133 Milano, Italy} %
\begin{abstract}
We address the use of a single qubit as a quantum probe to characterize
the properties of classical noise. In particular, we focus on the
characterization of classical noise arising from the interaction with a
stochastic field described by Gaussian processes.  The tools of quantum
estimation theory allow us to find the optimal {\em state preparation}
for the probe, the optimal {\em interaction time} with the external
noise, and the optimal {\em measurement} to effectively extract
information on the noise parameter.  We also perform a set of  simulated
experiments to assess the performances of maximum likelihood estimator,
showing that the asymptotic regime, where the estimator is unbiased and
efficient, is approximately achieved after few thousands repeated
measurements on the probe system.
\end{abstract}
\date{\today}
\end{frontmatter}
\section{Introduction}
Quantum systems of interest for quantum technology are usually immersed
in complex environments, which influence their dynamics and generally
induce decoherence. The characterization of the environment properties
is thus a relevant topic for the development of effective quantum
protocols.  In many situations, the environment may be conveniently
represented as a collection of fluctuators, such that it can be described
as a classical stochastic field, e.g. driven by a Gaussian process.
In fact, much attention has been recently devoted to answering the
question whether even a quantum bath can be described using a classical
or semi-classical picture of the environment 
\cite{yacoby11,bluhm11,DasSarma13,bluhm14,crow14}.
The classical description becomes progressively more reliable as far as 
the environment has many degrees of freedom or when the the interaction 
between a quantum system and a classical fluctuating field is taken into 
account. Several systems of interest indeed belong to this category, 
including the dynamics of quantum correlations in the presence of classical 
fluctuations \cite{eberly10,li11,colored,fluc,benedeIJ}, the simulation of 
motional averaging \cite{paraoanu14}, and the decoherence problem associated 
to the non-Markovian dynamics of solid state qubits 
\cite{bukard,bergli12,nonmark}.
\par
A reliable characterization of the environment, e.g. through 
its power spectrum, may allow one to design robust quantum protocols
resilient to noise \cite{bylander,zhang07,ahmed13,almog}. To this aim, some efforts have 
been recently devoted to understand whether the (de)coherent dynamics 
of a qubit can be used to extract information on the noise affecting
the qubit itself \cite{alvarez11,fbm14,benedettiNG,lucasz}. The canonical
way to attack this problem is by  using the tools of quantum estimation 
theory (QET) 
\cite{hel76,mal93,brau94,brody99,paris09,esch12}. Indeed, QET allows one to individuate the
best strategy to estimate the value of an unknown parameter, even when
it corresponds to a quantity which is not accessible by direct measurement. Upon
collecting the outcomes from the measurement of a suitably optimized
observable,  it is possible to build an estimator and infer the value of 
the parameter with the ultimate precision allowed by quantum mechanics. 
QET has been effectively employed in several scenarios, e.g 
to estimate quantum correlations \cite{brida,blandino}, interferometric 
phase-shift \cite{monras06,smerzi07,olivares09,teklu,kac10,cable10,
durk10,genoni2011,spagnolo2012}, and the spectral properties of non-Gaussian 
environments \cite{fbm14,benedettiNG}. Concerning quantum probes, optimized
quantum thermometry by single qubit has been recently addressed
experimentally \cite{mqt3,mqt4} and theoretically \cite{mqt1,mqt2}.
\par
In this paper we address the characterization of classical noise using 
a qubit as a quantum probe, and focus
attention to noise generated by Gaussian stochastic processes, i.e.
processes that are fully described by their power spectrum or their 
autocorrelation function. 
A relevant example of Gaussian process is the Ornstein-Uhlenbeck process,
which has been extensively employed in various contexts 
\cite{dechiara,mao,fiasconaro,benedettiou}. 
For the sake of completeness, and in order to analyze possible effects
due to specific features of the noise spectra, we also consider the
noise generated by processes with a Gaussian or a power law
autocorrelation function. 
\par
The performances of a qubit as a quantum probe clearly depend on the
kind of interaction it establishes with the environment. In order to
maintain the analysis self-contained, and to address situations of
practical interest, in the following we will assume that the dephasing
effects of the environment are much stronger than relaxation (damping)
ones. This generally happens when the typical frequencies of the
environment are smaller than the natural frequencies of the system, i.e.
the energy splitting between the eigenstates of the unperturbed
Hamiltonian. In this case, in fact, fluctuations can cause a
superposition to decohere, without driving transitions between the
different levels. In this framework, the characterization of 
the noise, which consists in estimating the parameters of the 
autocorrelation function, amounts to estimate the 
characteristic time describing the dephasing process occurring during
the decoherent dynamics.
\par
Due to the relatively simple dynamics of the probe, we have been able 
to evaluate the quantum Fisher information (QFI) and the quantum
signal-to-noise ratio (QSNR) analytically. Upon 
maximizing the QFI we obtain the three ingredients required 
to build an optimized
inference strategy to characterize the noise, i.e.: i) the 
optimal initial state preparation for the qubit; ii) the optimal 
interaction time with the environment; iii) the optimal measurement to
be performed at the output. The final step is then the processing 
of data to infer the value of the noise parameter, for which we employ 
a maximum likelihood estimator (MLE). In order to assess the performances of 
MLE we have performed a set of simulated experiments, showing that the 
asymptotic regime, where it becomes unbiased and efficient, is
approximately achieved after few thousands repeated measurements on the probe
system.
\par
The paper is organized as follows: in Sec. \ref{sec1} we introduce the
physical model for the qubit-environment system and describe the
Gaussian processes generating the noise; in Sec. \ref{sec2} we briefly
review the tools of local quantum estimation theory; in Sec. \ref{sec3}
we present our results about the optimal settings to achieve a large QFI
and the performances of a likelihood estimator. In Sec. \ref{sec4} we
end the paper with some concluding remarks.
\section{The physical model}\label{sec1}
Consider a qubit interacting with a classical fluctuating field which
induces dephasing. The qubit Hamiltonian is given by
\begin{equation}
 \mathcal{H}(t)=\omega_0\sigma_z+B(t)\sigma_z,\label{ham}
\end{equation}
where $\omega_0$ is the qubit energy, $\sigma_z$ is the Pauli matrix,
and $B(t)$ is a stochastic stationary process that follows a Gaussian statistics.
In particular, we focus on processes characterized by a zero mean and a
autocorrelation function $K(t,t')$, in formula:
\begin{align}
 &\expec{B(t)}{B}=0\\
 &\expec{B(t)B(t')}{B}=K(t-t')
\end{align}
where $\expec{...}{B}$ represents the average over the stochastic 
process $B$.
A Gaussian process is fully described by its second order statistics, 
e.g. its autocorrelation function $K$. The characteristic
function is given by \cite{wie24,wio13}:

\begin{align}
 \expec{
 e^{i\int_{t_0}^t ds\,f(s)\,B(s)}
 }{B}=
 e^{
 -\frac{1}{2}\int_{t_0}^t\int_{t_0}^t
 ds\,ds'\,f(s)K(s-s')f(s')}\,.\label{averC}
\end{align}
From the Hamiltonian \eqref{ham}, we can write the time evolution operator
\begin{align}
U(t)=\exp\left\{-i\int_0^t\mathcal{H}(s)ds\right\}=
\exp\left\{
-i[\omega_0t+\varphi(t)]\sigma_z
\right\}\,,
\end{align}
where we defined the noise phase {$\varphi(t)=\int_0^t B(s)\,ds$}.
We assume that the qubit is initially in a pure state
$\ket{\psi_0}=\cos\theta/2\,\ket{0}+\sin\theta/2\, \ket{1}$
with $0<\theta<\pi$. The qubit density matrix is given by 
the average of the evolved density matrix 
over the stochastic process:
\begin{align}
 &\rho(t)=\expec{U(t)\rho(0)U^{\dagger}(t)}{B}=\nonumber\\
 \vspace{50pt}
 &\frac{1}{2}\left(\begin{array}{cc}
          1+\cos\theta&e^{-2i\omega_0t}\expec{e^{-2i\varphi(t)}}{B}\sin\theta\\
          e^{2i\omega_0t}\expec{e^{2i\varphi(t)}}{B}\sin\theta&1-\cos\theta
         \end{array}
\right),\label{rhot}
\end{align}
where the initial state is $\rho(0)=\ketbra{\psi_0}{\psi_0}$.
We can rewrite Eq. \eqref{rhot} as:
\begin{equation}
 \rho(t)=\frac{1}{2}\left(\begin{array}{cc}
                1+\cos\theta&e^{-2(i\omega_0 t+\beta(t))}\sin\theta\\
                \\
                e^{2(i\omega_0 t-\beta(t))}\sin\theta&1-\cos\theta
               \end{array}\right),\label{rhot2}
\end{equation}
where the off diagonal terms are calculated using Eq. \eqref{averC} and
the function $\beta$ is related to the autocorrelation function of the stochastic
process generating the classical noise through the relation:
\begin{align}
 \beta(t)=\int_0^t\!\int_0^t\!\!ds\,ds'\,K(s-s')\label{beta}.
\end{align}
In this paper we consider three particular Gaussian processes. Specifically,
we assume that the stochastic field $B(t)$ in Eq. \eqref{ham} is driven
either by an Ornstein-Uhlenbeck (OU), or by a Gaussian (G) or power-law
(PL) one. The corresponding autocorrelation functions are given by
\begin{align}
 K_{OU}(t-t', \gamma,\Gamma)&=\frac{ \Gamma\gamma}{2}\, \,e^{-\gamma|t-t'|}\quad\\
 K_{G}(t-t',\gamma,\Gamma)&=\frac{\Gamma\gamma}{\sqrt{\pi}}\,e^{-\gamma^2(t-t')^2}\\
 K_{PL}(t-t',\gamma,\Gamma,\alpha)&=\frac{\alpha-1}{2}\,
\frac{ \gamma\Gamma}{\big(\gamma|t-t'|+1\big)^{\alpha}}\label{plc}
\end{align}
where $\gamma$ is the unknown noise parameter, $\Gamma$ is the damping
rate that we assume fixed, and $t$ is the interaction time. In Eq. (\ref{plc}) we have the
constraint $\alpha >2$.
Inserting these autocorrelation functions in Eq. \eqref{beta} leads to the 
following $\beta$ functions:
\begin{align}
 \beta_{OU}(g,\tau)&=\frac{1}{g}\left(g\tau+e^{-g\tau}-1\right)\label{bou}\\
 \beta_G(g,\tau)&=\frac{1}{g}\left[g\tau\,\hbox{Erf}(g\tau)+
 \frac{e^{-(g\tau)^2}-1}{\sqrt{\pi}}\right]\label{bg}\\
 \beta_{PL}(g,\tau)&=\frac{1}{g}\,\left[\frac{(1+g\tau)^{2-\alpha}+g\tau(\alpha-2)-1}{(\alpha-2)}\right]\label{bpl}.
\end{align}
where we introduced the adimensional quantities 
$g=\frac{\gamma}{\Gamma}$ and $\tau=\Gamma t$.
\par
The characterization of the classical noise amounts to estimate
the overall noise parameter $g$ by performing 
measurements on the quantum probe after the interaction, i.e. on the
states described by the density matrices in Eq. (\ref{rhot2}).
In order to make this procedure as effective as possible, i.e. to
extract the maximum amount of information on the noise by inspecting the
state of the probe, we have to
suitably optimize the initial preparation of the qubit, the value of the 
interaction time, the measurement to be performed at the output and,
finally, the data processing after collecting an experimental sample.
The proper framework to attack this optimization problem is that of 
local quantum estimation 
theory  \cite{hel76,mal93,brau94,brody99,paris09,esch12},
which we are going to briefly
review in the next Section. 
\section{Quantum estimation theory}\label{sec2}
Consider a family of quantum states $\rho_{\gamma}$, characterized
by an unknown value of a parameter $\gamma$, usually corresponding 
to a non-observable quantity. 
The goal of any estimation procedure is to infer the
value of the unknown parameter $\gamma$ by measuring some observable 
quantity on the system $\rho_{\gamma}$. This is achieved by 
collecting the outputs $(x_1,x_2,\dots,x_M)$ of such measurements and
use them to build an estimator $\hat{\gamma}=\hat{\gamma}(x_1,x_2,\dots,x_M)$,
i.e. a function of the outcomes. The smaller is the estimator variance 
$\sigma^2$ (over data), the more
accurate is the estimation procedure. The lower bound to the precision of any
unbiased estimator is given by the Cram\'er-Rao (CR) bound:
\begin{equation}
\sigma^2(\hat{\gamma})\geq\frac{1}{M\,F(\gamma)},
\label{ccr}
\end{equation}
where $M$ is the number of measurements and $F(\gamma)$ is the Fisher
Information (FI):
\begin{align}
 F(\gamma)=\sum_x \, p(x|\gamma)\left[\partial_{\gamma}
 \ln\,p(x|\gamma)\right]^2\,,
 \label{fi}
\end{align}
where $p(x|\gamma)$ is the conditional probability of obtaining the outcomes $x$
if the true value of the parameter is $\gamma$. Given a quantum system, 
the conditional probability can be written as $p(x|\gamma)=\hbox{Tr}[\rho_{\gamma}E_{x}]$
with $E_x$ a positive operator-valued measure (POVM). By maximizing the 
FI over all possible POVMs (see e.g. \cite{paris09}), one obtains the 
ultimate bound to the precision of any estimator, i.e.
the quantum Cramer-Rao (QCR) bound:
\begin{equation}
 \sigma^2(\hat{\gamma})\geq\frac{1}{M\,H(\gamma)},\label{qkr}
\end{equation}
where $H(\gamma)$ is the quantum
Fisher information, i.e. the superior of $F(\gamma)$ over POVMs. 
A measurement $E_x$ is said to be \emph{optimal} when its FI coincides with the QFI, i.e. 
$F(\gamma)=H(\gamma)$.
Eqs. \eqref{ccr} and \eqref{qkr} set the lower bound to the
precision of any estimation procedure. Once a measurement has been
chosen, and performed, one has to process data, i.e. to choose an
estimator. Estimators for which the CR bound is saturated are said to be 
\emph{efficient}. 
\par
For a family of qubit states, the QFI reads:
\begin{align}
 H(\gamma)=\sum_{n=1}^{2}\frac{(\partial_{\gamma}p_n)^2}{p_n}+2\sum_{n\neq m}
 \frac{(p_n-p_m)^2}{p_n+p_m}|\braket{p_m}{\partial_{\gamma}p_n}|^2
\label{qfi}
 \end{align}
where $p_n$ and $\ket{p_n}$ are respectively the 
eigenvalues and eigenvectors of the qubit density matrix
$\rho=\sum_{n=1,2}p_n\ketbra{p_n}{p_n}$.
\par
A suitable figure of merit to assess the overall estimability of a parameter
is the quantum signal-to-noise ratio (QSNR):
\begin{equation}
 R(\gamma)=\gamma^2H(\gamma),\label{snr}
\end{equation}
which accounts for the fact that large values of the parameter are
generally easier to estimate, while small values need more 
precise estimators.
A given parameter is said to be easily estimable if the corresponding $R$ 
is large. On the other hand, if $R$ is small the estimation of $\gamma$
is an inherently inefficient procedure, whatever strategy is employed to
infer its value. 
\par
Once a measurement has been chosen, possibly the optimal one maximizing
the Fisher information, one has to chose an estimator, i.e. a procedure
to process data in order to infer the value of the parameter of
interest. An estimator which is {\em asymptotically} efficient, i.e.
it saturates the QCR bound in the limit of large samples, is the 
maximum likelihood estimator. Consider $M$ independent 
measurements of the random variables $X$, with probability density $p(x|\gamma)$.
The joint probability function of an experimental sample of size $M$, 
$\{x_i\}_{i=1}^M$, is given by the product $\prod p(x_i,\gamma)$, and it
is usually referred to as the Likelihood  function $L(\gamma)$ 
\begin{align}
L(\gamma)= L(\gamma|x_1,x_2,\dots,x_M)=\prod_{i=1}^Mp(x_i|\gamma).\label{likf}
\end{align}
The MLE for the parameter $\gamma$ is the value yielding the largest 
likelihood of the observed values, that is the value that maximizes 
the quantity 
in Eq. \eqref{likf}:
\begin{align}
\hat{\gamma}_\ml =\arg\max_{\gamma} L(\gamma)\,.\label{gml}
\end{align}
As mentioned above, $\gamma_\ml$ is known to be asymptotically
efficient \cite{leh98}, i.e. it saturates the CR bound for large number of
measurements $M\gg 1$. On the other hand, in practical situations, 
one is usually interested in checking whether this regime is achieved
for values of $M$ within the experimental capabilities.
\section{Quantum probes for classical environments}\label{sec3}
In this section we present and discuss our results. In the first
subsection, we find the analytic expressions of the QFI and the QSNR
for the estimation of the noise parameter $g$ of the considered processes. 
Moreover, we show that the optimal
measurement corresponds to the Pauli matrix $\sigma_x$ in the rotating
frame of the qubit.
In the second subsection we
assess the performances of the MLE by a set of 
simulated experiments. 
\subsection{Signal-to-noise ratio and optimal setting}
The QFI gives the ultimate quantum bound to the precision of 
an inference procedure. For the family of qubit density 
matrices described by Eq. \eqref{rhot2}, the QFI can be computed 
using Eq. \eqref{qfi},
through the eigenvalues and eigenvectors of the density operator:
\begin{align}
&p_{\pm}(g,\tau)=\frac{1}{2}\left(1\pm e^{-2\beta(g,\tau)}
\right)\label{eigv}\\
&\ket{p_{\pm}}=\frac{1}{\sqrt{2}}\left(\pm e^{-2i\omega_0t}
\ket{0}+\ket{1}\right).\label{autov}
\end{align}
where we substituted the symbol $p_{1,2}$ with $p_{\pm}$ 
to denote eigenvalues and eigenvectors. Inserting these 
expressions in Eq. \eqref{qfi}, one obtains the analytic 
expression for the QFI:
\begin{align}
 H(g,\tau)=4\frac{\sin^2\theta}
 {e^{4\,\beta(g,\tau)}-1}\,
 \left[\partial_{g}\, \beta(g,\tau)\right]^2
 \,.\label{qfi1}
\end{align}
It is worth noticing that Eq. \eqref{qfi1} does not depend on 
the qubit energy $\omega_0$, and it is maximized by 
$\theta=\frac{\pi}{2}$. It follows that the optimal initial state
is the superposition 
$\ket{\psi_0}=\frac{1}{\sqrt{2}}(\ket{0}+\ket{1})=\ket{+}$.
\begin{figure}[h!]
\begin{tabular}{cc}
\includegraphics[width=0.47\columnwidth]{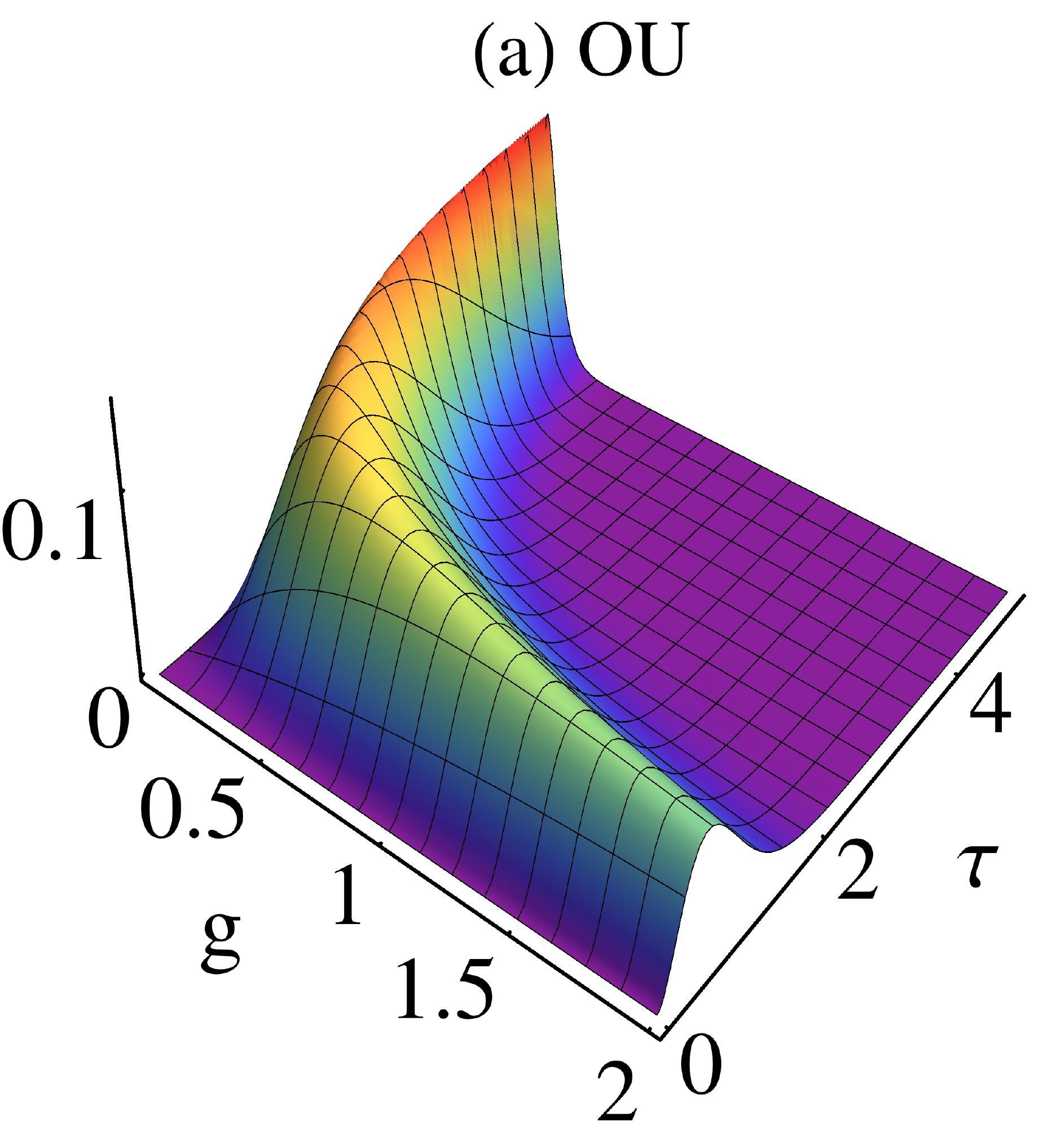}&
\includegraphics[width=0.47\columnwidth]{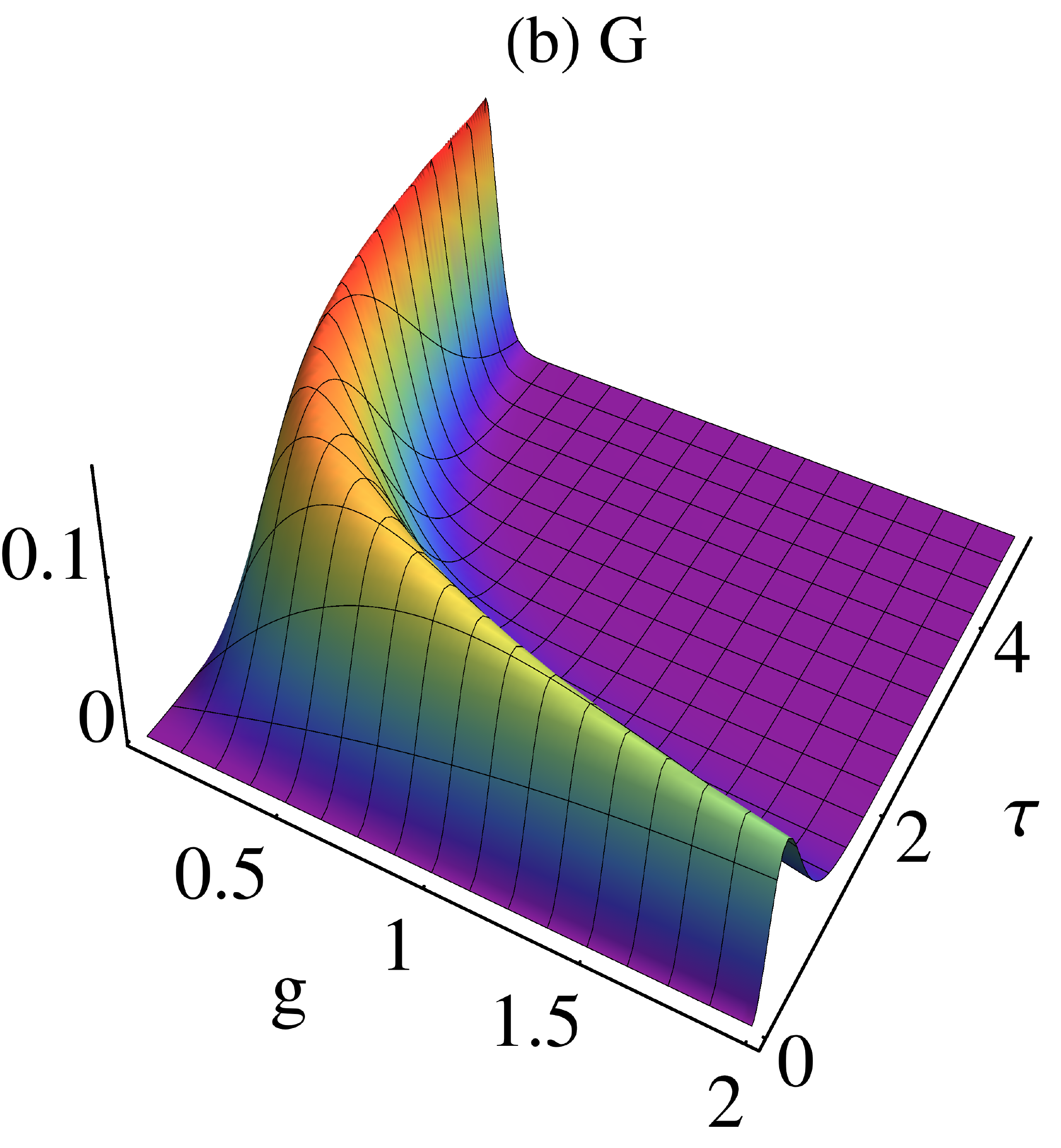}\\
\includegraphics[width=0.47\columnwidth]{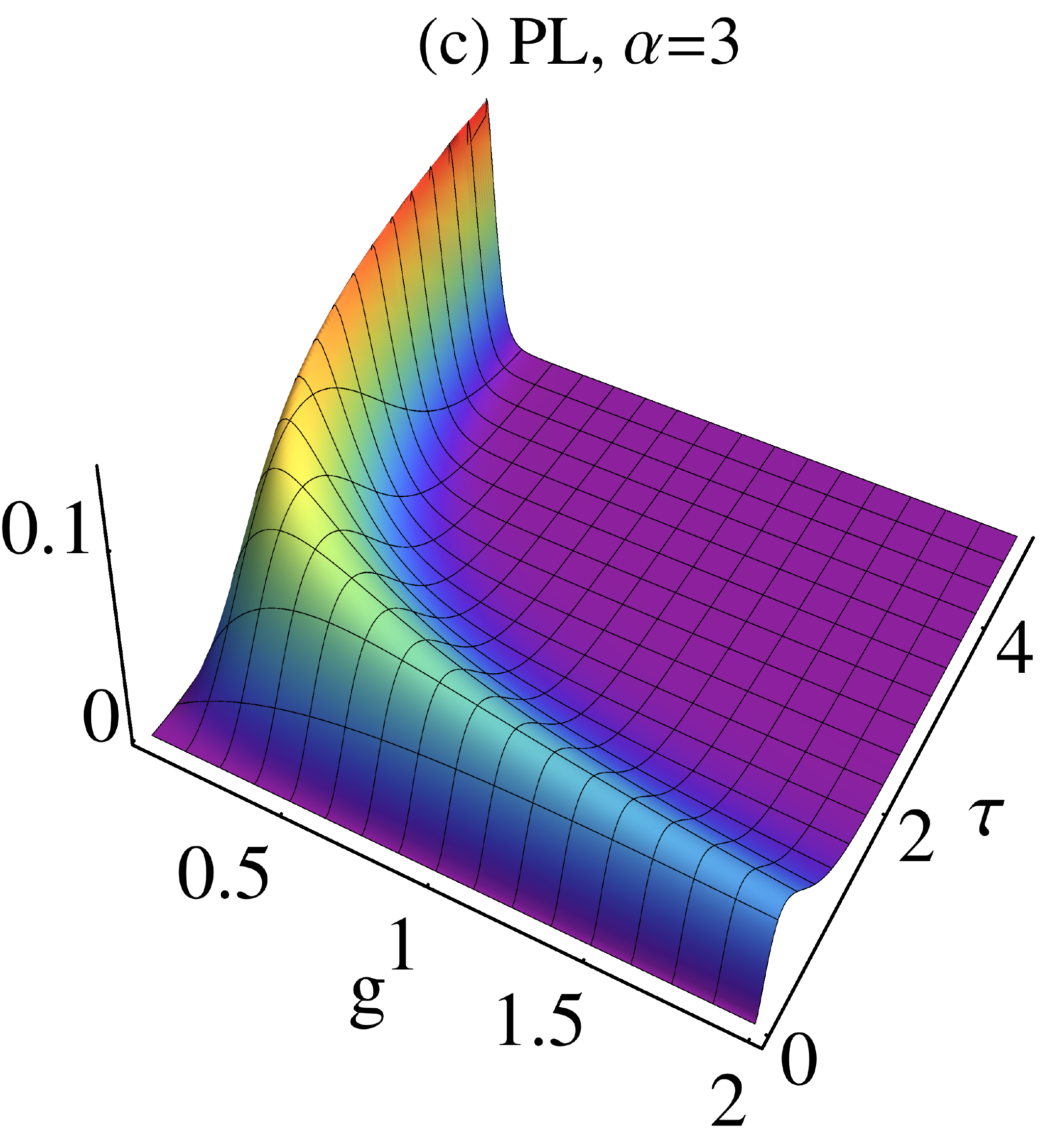}&
\includegraphics[width=0.47\columnwidth]{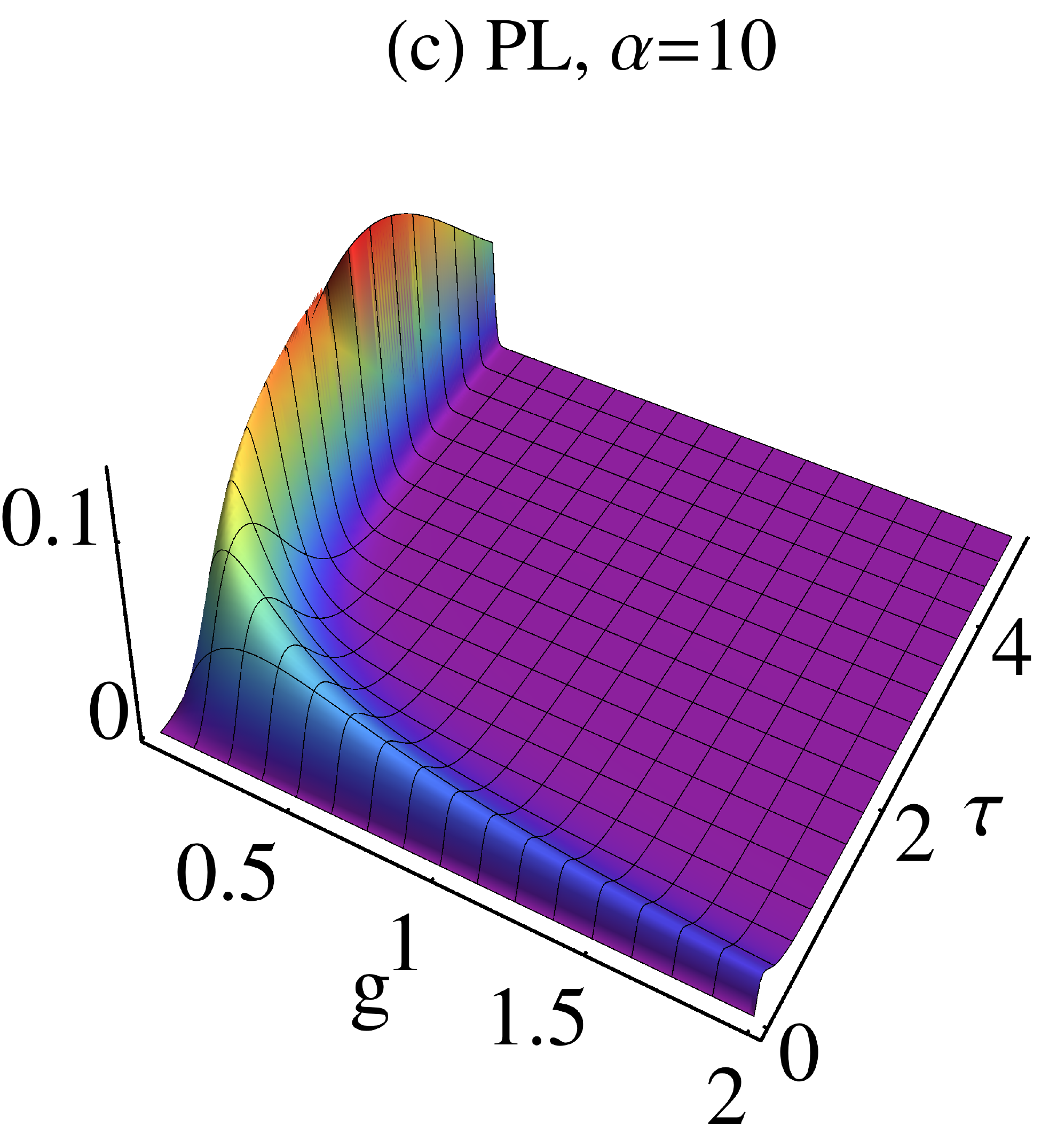}
\end{tabular}
\caption{(Color online): The quantum signal-to-noise ratio $R(g)$ as a function 
of $g$ and the interaction time $\tau$ for different stochastic processes: (a) OU,
(b) G, and PL with (c) $\alpha=3$ and (d) $\alpha=10$.}\label{fig1}
\end{figure}
\par
\begin{figure}[h!]
\centering
\includegraphics[width=0.9\columnwidth]{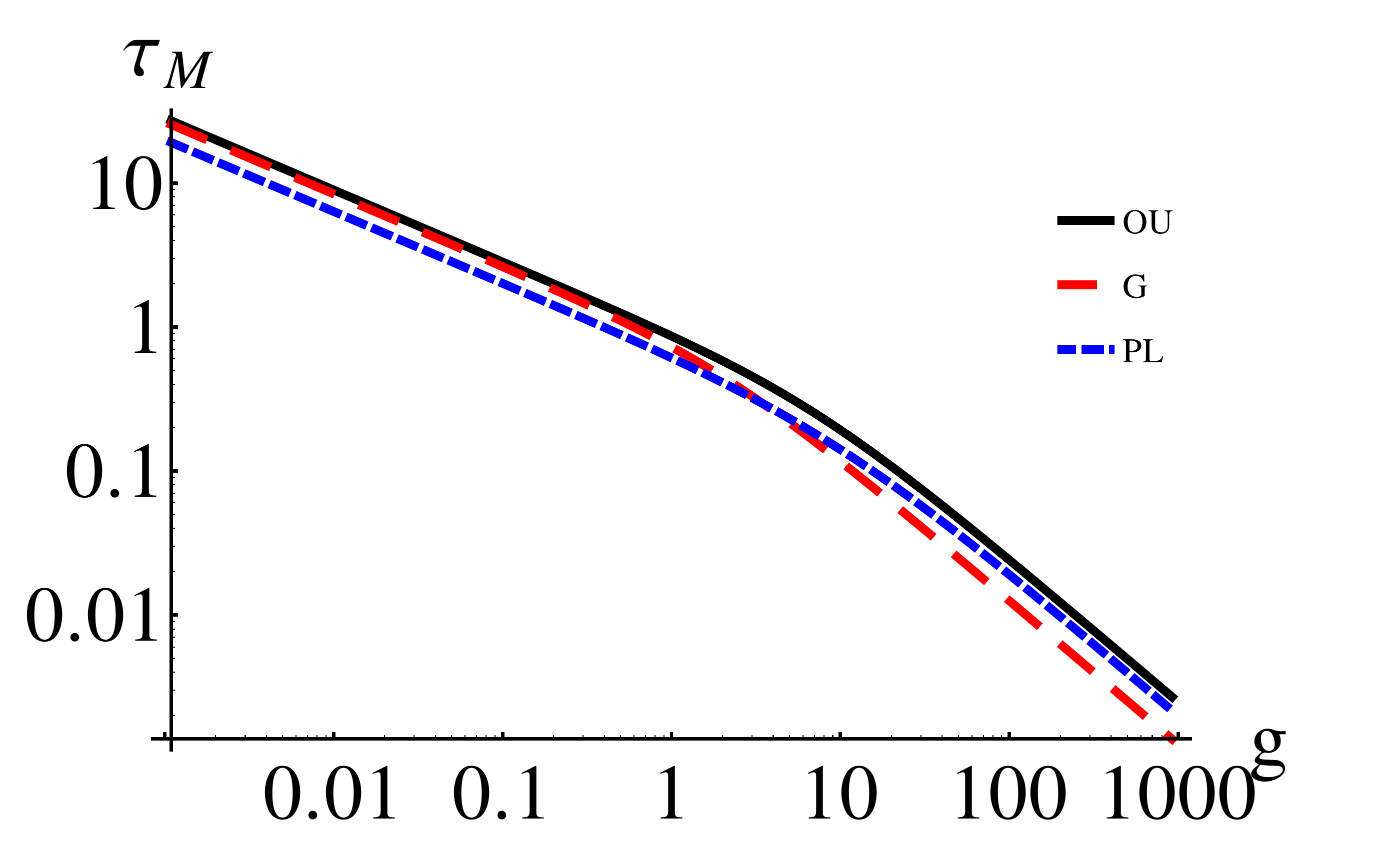}
\includegraphics[width=0.9\columnwidth]{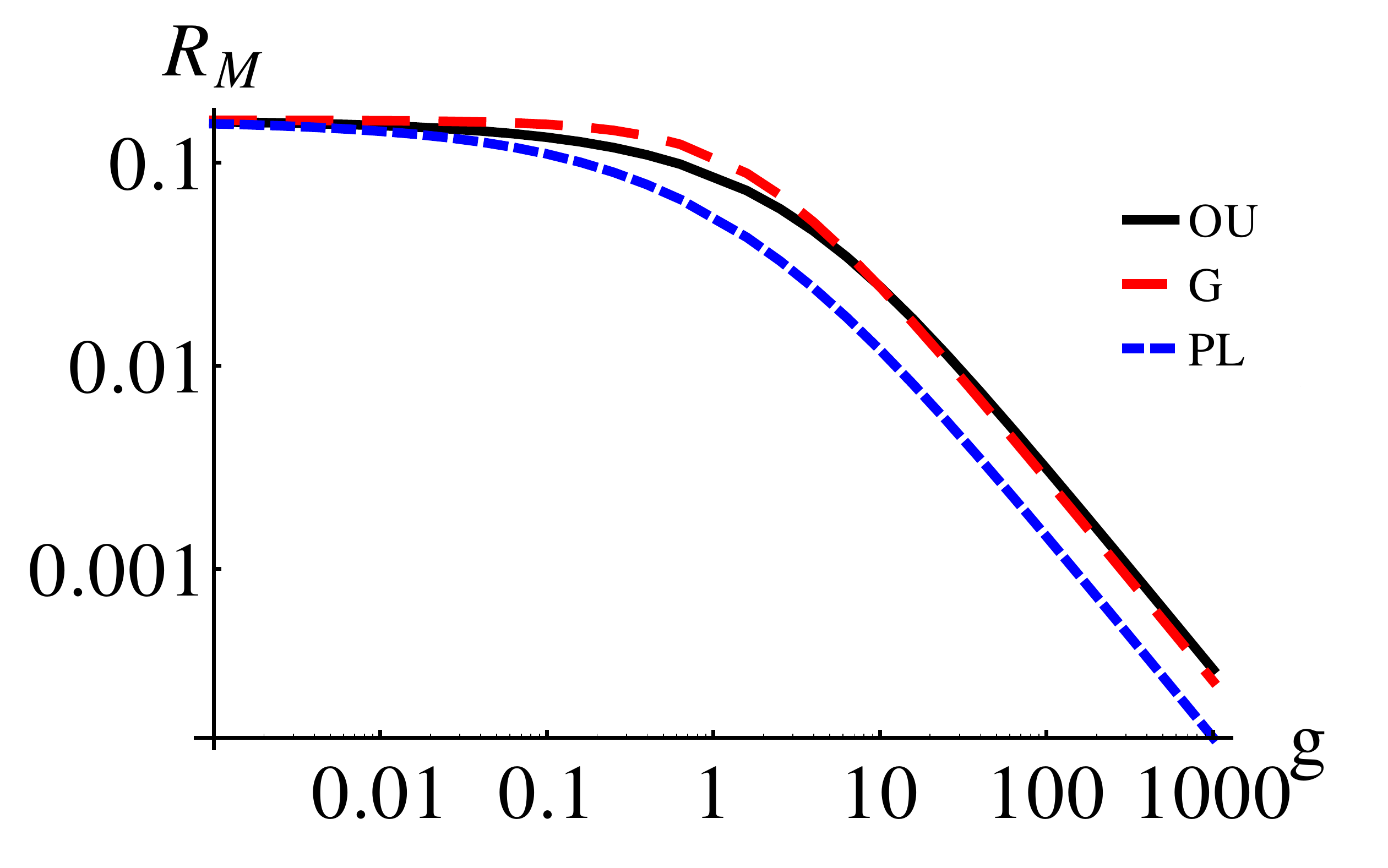}
\caption{(Color online): The upper panel shows the optimal interaction 
time $\tau_\opt (g)$, which maximizes the QSNR, for the three different
processes. We have OU (solid black), G (dashed red) and PL (dotted blue). 
In the PL case, we set $\alpha=3$. The lower panel shows the
corresponding (maximized) values of the QSNR $R_\mx$, using the same
color code.
}\label{f2res}
\end{figure}

For the processes described in Eqs. \eqref{bou}-\eqref{bpl}, the 
QSNR is calculated
from Eq. \eqref{snr} and it is given by:
\begin{align}
&R_{OU}(g,\tau)=\frac{4\,e^{-2g\tau}}{g^2}\left[\frac{(1-e^{g\tau}+g\tau)^2}
  {e^{4\left(\tau+\frac{e^{-g\tau}-1}{g}\right)}-1}\right]
  \label{hou}\\
 &R_{G}(g,\tau)=\frac{4}{\pi g^2}\left[\frac{\left(e^{-g^2\tau^2}-1\right)^2}
  {e^{4 \left(\frac{e^{-g^2\tau^2}-1}{\sqrt{\pi}g}+
  \tau\,\text{Erf}(g\tau)\right)}-1}\right]\nonumber\\
&R_{PL}(g,\tau)=\nonumber\\
&\frac{4}{g^2}\left[\frac{(1+\alpha g\tau+(\alpha-1)(g\tau)^2-(1+g
\tau)^{\alpha})^2} {\left(e^{4\left(\tau+\frac{(1+g\tau)^{2-\alpha}
-1}{g(\alpha-2)}\right)-1}\right)(\alpha-2)^2(1+g\tau)^{2\alpha}}\right]
\nonumber
\end{align}
The QSNRs of Eqs. (\ref{hou}) are shown in Fig. \ref{fig1}. As it is 
apparent from the plots, 
the qualitative behavior is the 
same for all processes. At any fixed value of $g$ there is 
a maximum in the QSNR, achieved for an optimal value of the interaction 
time $\tau_\opt (g)$. The value of this maximum $R_\mx =
R(\tau_\opt)$ decreases with $g$. It follows that smaller values 
of $g$ may be better estimated than larger ones. 
The optimal time $\tau_\opt (g)$ decreases with increasing
values of the parameter. This means that the  smaller is $g$, the longer 
is the interaction time that is required to effectively {\em imprint} the 
effects of the external environment on the probe.
The dependency of $\tau_\opt$ on the parameter $g$ is shown in 
the upper panel of Fig. \ref{f2res}, for the three considered 
processes. For small values 
of $g$ we have approximately $\tau_\opt \simeq a/\sqrt{g}$ (with
$a\simeq 0.89$ for OU and similar values for the other processes) while for
$g\gg 1$ we may write $\tau_\opt \simeq b/g$, with $b\simeq 2.5$ for OU.
The corresponding values of the QSNR, i.e. $R_\mx$ are shown in the
lower panel of the same figure. $R_\mx$ is almost constant for small
$g$ and then start to decrease. We have $R_\mx \simeq a - b \sqrt{g}$
for $g\ll1$, where $a\simeq 0.161$ and $b=0.096$ for OU, 
and $R_\mx \simeq b/g$ for $g\gg 1$, with $b\simeq 0.33$ for OU.
It follows that $g$ may be effectively estimated when it is small, since
the QSNR is large. In this regime, the estimation procedure is also robust, 
since the optimal interaction time and the resulting value of the QSNR 
depend only weakly on the value of $g$. On the other hand, for larger
$g$ the estimation procedure is unavoidably less effective.
\par
To complete our analysis, we now prove that the optimal measurement
achieving the QFI is a realistic one, since it corresponds to the projectors 
onto the eigenstates \eqref{autov}. In fact, the FI of the distributions 
\eqref{eigv}, computed from Eq. \eqref{fi}, is given by:
\begin{align}
 F(g,\tau)&=\frac{[\partial_{g}p_{+}(g,\tau)]^2}{p_{+}(g,\tau)}+
 \frac{[\partial_{g}p_{-}(g,\tau)]^2}{p_{-}(g,\tau)}\nnum
 &=\frac{4 \;
 [\partial_{g}\beta(g,\tau,t)]^2}{e^{4\beta(g,\tau)}-1}=H(g,\tau)\,,
 \end{align}
which coincides with the QFI.
The optimal measurement is thus obtained from the projectors 
onto the eigenstates of the density
matrix $\Pi_{\pm}=\ketbra{p_{\pm}}{p_{\pm}}$:
\begin{align}
 \Pi_{\pm}&=\frac{1}{2}\left(\begin{array}{cc}
                 1&\pm e^{-2i\omega_o t}\\
                 \pm e^{2i\omega_o t}&1
                \end{array}\right)\\
           &= \frac{1}{2}e^{-i\omega_0 t\sigma_z}
	   \ketbra{\pm}{\pm}e^{i\omega_0 t\sigma_z}.\label{ppm}
\end{align}
In other words, the optimal measurement corresponds to $\sigma_x$ in the
qubit reference frame which rotates with frequency $\omega_0$. 
\subsection{Maximum Likelihood estimator}
In this Section we present the results of simulated
experiments, performed to assess the performances of the MLE 
and to characterize its asymptotic regime.
In particular, we have numerically simulated repeated measurements of
the observable described by the projectors $\Pi_\pm$ in Eq. 
\eqref{ppm},  and then estimated the value 
of the parameter $g$ in the case of OU process using MLE.
\par
Let us consider to have performed $M$ repeated measurements 
of $\Pi_{\pm}$ at the optimal time $\tau_\opt$
Each run returns $\pm1$, according 
the probability distributions \eqref{eigv}. Let us call $N$ the number of
outcomes with value $+1$. The frequentist interpretation of probability leads us to
write the relation 
\begin{align}
 p_+(g,\tau)=\frac{N}{M}\,,\label{freqp}
\end{align}
 implicitly assuming that the number of measurement is large $M\gg1$. 
\begin{figure}[h!]
\includegraphics[width=0.9\columnwidth]{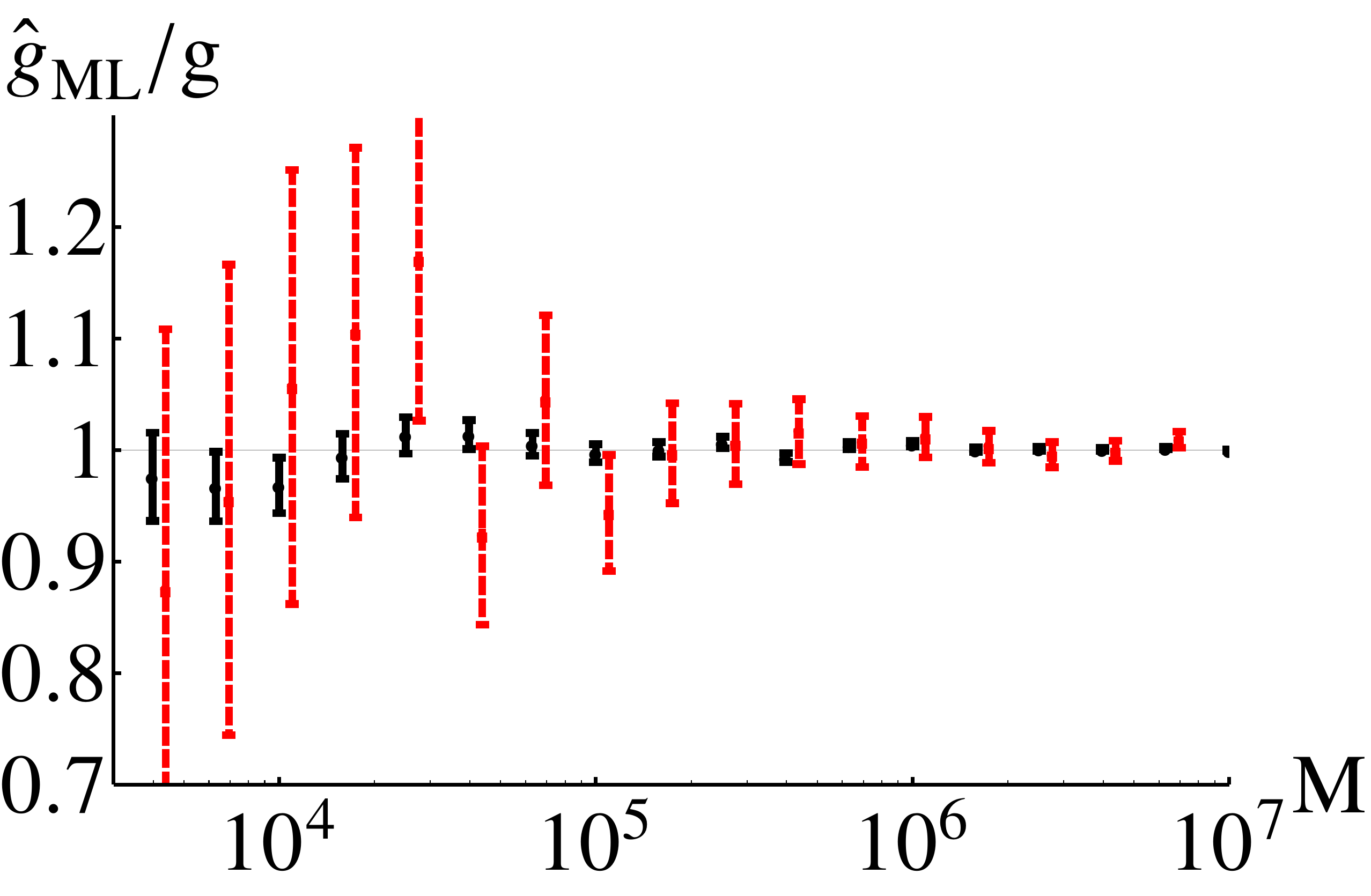}\\
\includegraphics[width=0.9\columnwidth]{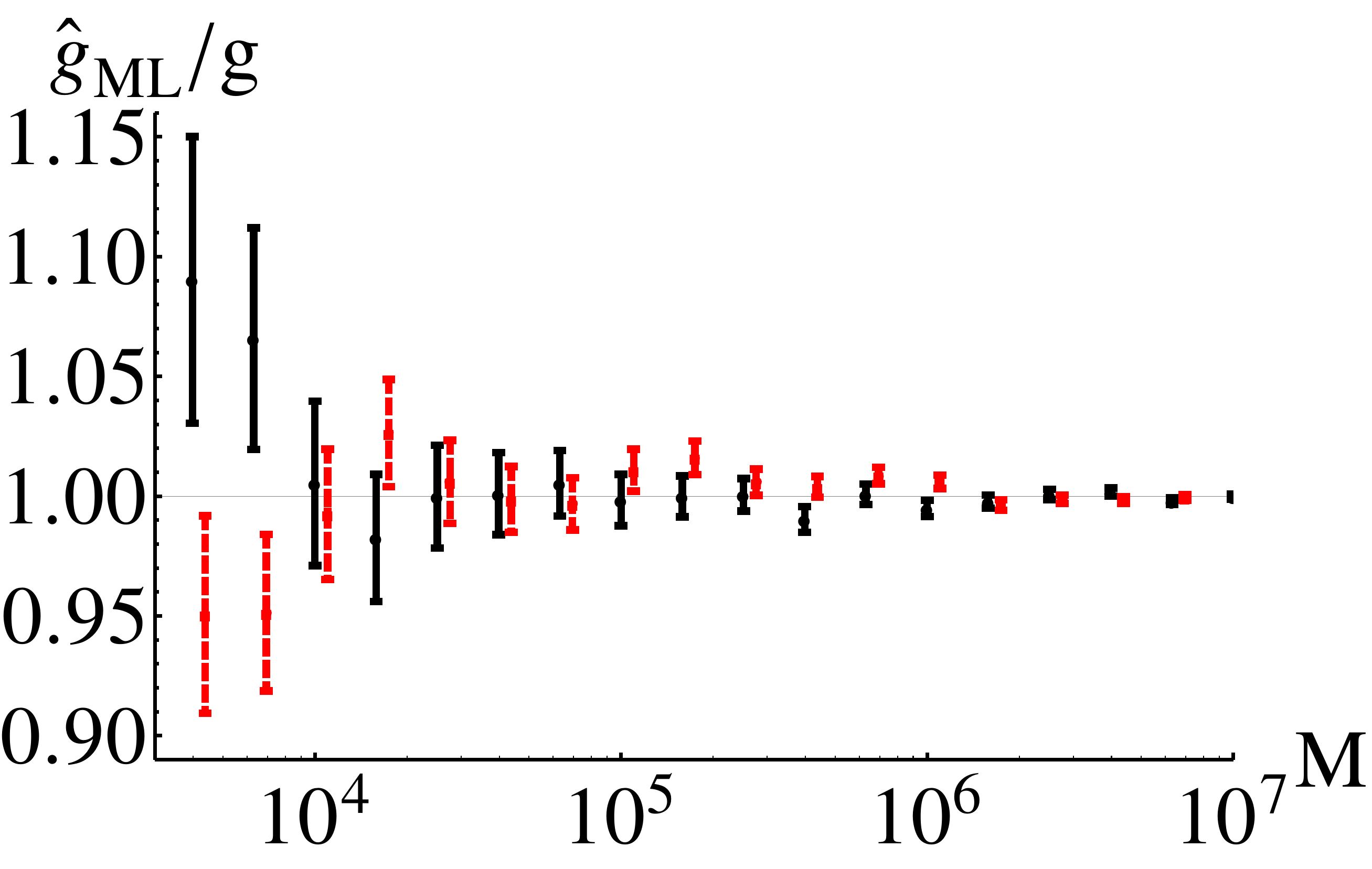}
\caption{(Color online): 
The two panels show the ratio $g_{\ml}/g$ between the ML estimated
value of $g$ and the true value, together with the corresponding error 
bars, as a function of the number of repeated (simulated) measurements 
$M$. In the upper panel the results for the true values $g=0.01$ 
(solid black) and $g=100$ (red dashed) are compared. Larger
values of the parameter are better estimated. In the lower panel the 
considered values are $g=0.1$ (solid black) and $g=1$ (red dashed).
Notice that the simulated data in both panels are computed for the 
same values of $M$ and then the red points are slightly shifted 
along the $x$-axes for the sake of clarity.}\label{fig2}
\end{figure}
\par
In order to simplify the notation, we hereafter call
$p(g,\tau)\equiv p_+(g,\tau)$. By inverting Eq. \eqref{freqp}, we can write the 
inversion estimator $\hat{g}$ of $g$: $\hat{g}(N,M)=p^{-1}\left(\frac{N}{M},\tau\right)$.
Before analyzing the performances of this estimator we show that
it coincides with the MLE. In fact, from Eqs. \eqref{likf}
and \eqref{gml}
we have:
\begin{align}
 P_L(g,\tau)=&p(g,\tau)^{N}[1-p(g,\tau)]^{M-N}\\
 \partial_{g}P_L(g,\tau)=&-[1-p(g,\tau)]^{M-N-1}p(g,\tau)^{N-1}\nonumber\\
 & \times [M p(g,\tau)-N]\,\partial_{g}p(g,\tau)\label{dml}.
\end{align}
Eq. \eqref{dml} has a maximum for  $p(g,\tau)=\frac{N}{M}$ which, by inversion,
gives the inversion estimator
\begin{equation}
 \hat{g}_{\ml}(N,M)=p^{-1}\left(\frac{N}{M},\tau\right).\label{MLE}
\end{equation}
The MLE is a function of the number of repeated measurements $M$ and 
the number of outcomes with value $+1$, $N$. By numerical simulations, we 
mimic the results of experiments. The variance of the MLE  \eqref{MLE}
is computed using the error propagation theory. Upon assuming that the measure outcomes
follow a binomial distribution, the estimator variance $\sigma^2$is given by:
\begin{align}
 \sigma^2(\hat{g}_{\ml})=\left|\frac{\partial
 \hat{g}_{\ml}(N,M)}{\partial {N}}\right|^2N\left(1-\frac{N}{M}\right).
 \label{sml}
\end{align}
In Fig. \ref{fig2} we shows the ratio between the estimated value
$\hat{g}_{\ml}$ and the true value as a function of the number of
repeated measurements for different values of the true parameter $g$.
The estimated value oscillates around the true one, with standard
deviations $\sigma$ decreasing as a function of $M$. In fact, as the
number of measurements becomes larger, the ratio $\hat{g}_{\ml}/g$  gets
closer to unity. The error associated to each point is smaller with
increasing number of measurements.  The sets of data in Fig. \ref{fig2}
refer to $g=0.01$ (black solid line) and $g=100$ (red dashed line) in
the upper panel and $g=0.1$ (black solid line) and $g=1$ (red dashed line) 
in the lower one.  The upper panel in Fig. \ref{fig2} highlights the fact that
for the data associated to small $g$, the ratio converges more rapidly
to unity and with smaller error with respect to the case $g=100$.  This
is in agreement with the results of the previous subsection, where we
found that $R_\opt$ is larger for smaller values of the
parameter, meaning that the parameter is better estimable in the regime
$g\ll1$. The lower panel of Fig. \eqref{fig2} confirm the behavior
found in Fig. \ref{f2res}: in the region $g<1$ it is possible to easily
estimate the parameter almost independently on the value of $g$.\par As
already mentioned, the variance $\sigma^2$ decreases 
with increasing $M$. 
This is expected from the QCR bound in Eq. \eqref{qkr} because the 
QFI is a fixed quantity for fixed $g$, so the minimum error scales as $\frac{1}{M}$.
\begin{figure}[h!]
\includegraphics[width=0.9\columnwidth]{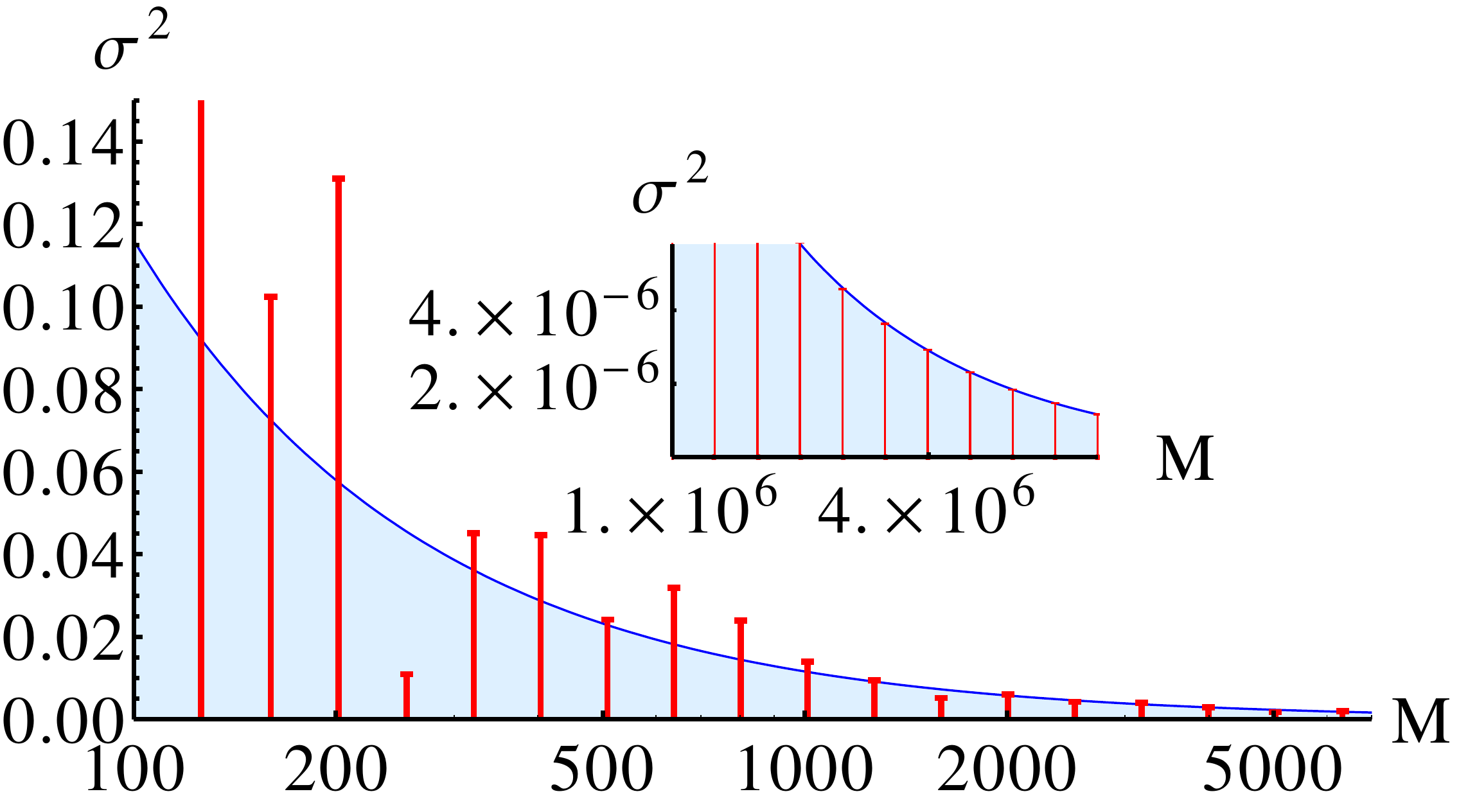}
\caption{(Color online): The variance (red line) 
of the ML estimator  as a function of  the number of measurements 
for the case $g=1$. The light blue area illustrates the QCR bound. 
Variances below the quantum bound mean that the estimator has a bias. 
Inset: The same as in the main frame but for a large number of 
measurements: the bias is no longer present.}\label{fig3}
\end{figure}
\par
In Fig. \ref{fig3} we illustrate the behavior of the variance $\sigma^2$
as a function of the measurement number in the case $g=1$. The red line 
represents the variance and the shaded area outlines the QCR bound. The
reader may note that in certain cases the variance is below the quantum
bound. This means that the estimator is slightly biased. But as the
number of measurements in increased the bias tends to zero and the
estimator becomes efficient (i.e. it saturates the QCR bound, as shown in the
inset) as expected for MLE.  The same qualitative behavior is
found for all the other values of the parameter $g$. 
From our analysis, we see that the asymptotic regime for MLE is already
achieved for a number of repeated measurements of about $10^4 - 10^5$. We have
also analyzed the convergence of a Bayesian estimator and found that the
required $M$ to have the asymptotic behavior is larger. It follows
that, to achieve the characterization of the spectral properties of a
Gaussian noise, a ML procedure lead to a faster estimation of the
unknown parameters.
\section{Conclusions}\label{sec4}
A detailed description of decoherence is crucial for 
the development of quantum information processing in 
realistic scenarios. In particular, the precise characterization 
of the noise acting on a quantum system is the main
tool in designing protocols to contrast its detrimental effects. 
In this paper, we have addressed the estimation of the 
noise parameter for Gaussian processes by the use a simple 
quantum system, such as a qubit, as a quantum probe. 
More specifically, by maximizing the quantum signal-to-noise ratio 
we have found the optimal setting to perform optimal measurements 
and inference.  Our results show that for
any fixed value of the estimable parameter, the QSNR has a maximum,
corresponding to an optimal value of the interaction time $\tau_\opt$.  
This maximum is larger for smaller values of the parameter, which 
may be estimated more precisely. 
\par
The ultimate bound to precision may be practically achieved by 
measuring the "polarization" of the qubit, i.e. the observable 
$\sigma_x$ in the rotating frame of the qubit, and then employing 
a maximum-likelihood estimator, which achieves the asymptotic 
regime, and thus the optimal performances, alredy after 
few thousands measurements. 
\par
{The estimation scheme presented in this paper would be
suitable also to infer the amplitude of white noise, characterized 
by an autocorrelation function $K$ proportional to a Dirac delta. In this
case, the optimal state preparation and measurement are the same 
as those obtained for Gaussian noise, However, the 
quantum signal-to-noise-ratio is a monotonically decreasing 
function of time, leaving no room for any optimization procedure.}
\par
At present, we cannot provide a quantitative
statement about the performance of quantum probes compared to
classical ones since the modelling of the latter would be rather
challenging. On the other hand, our results show that quantum 
probes, besides having the advantage of introducing small 
perturbations into the
system, require only measurements performed at a single instant
of time, thus avoiding the need of observing the system for a long time 
in order to collect a time series.
\section*{Acknowledgments}
This work has been supported by the MIUR project FIRB-LiCHIS-RBFR10YQ3H.  
The authors thank the FIM Department of University of Modena and
Reggio Emilia for hospitality and an anonymous referee for useful
suggestions.


\section*{References}
\begin{thebibliography}{99}
\bibitem{yacoby11} I. Neder, M. S. Rudner, H. Bluhm, S. Foletti, 
B. I. Halperin, A. Yacoby, Phys. Rev. B, {\bf 84}, 035441 (2011).
\bibitem{bluhm11} M. J. Biercuck, H. Bluhm, Phys. Rev. B {\bf 83}, 235316 (2011).
\bibitem{DasSarma13} W. M. Witzel, K. Young and S. Das Sarma, ArXix:1307.2597v1.
\bibitem{bluhm14} T. Fink, H. Bluhm, ArXiv:1402.0235v1.
\bibitem{crow14} D. Crow, R. Joynt, Phys. Rev. A {\bf 89}, 042123
(2014).
\bibitem{eberly10} T. Yu, J. H. Eberly, Opt. Comm. {\bf 283} (2010) 676.
\bibitem{li11} J.-Q. Li and J.-Q. Liang, Phys. Lett. A, {\bf 375}, 1496 (2011).
\bibitem{colored} C. Benedetti, F. Buscemi, P. Bordone, 
M. G. A. Paris, Phys. Rev. A {\bf 87}, 052328 (2013).
\bibitem{fluc} C. Benedetti, F. Buscemi, P. Bordone and 
M. G. A. Paris, Int. J. Quantum Inform. {\bf 10}, 1241005 (2012).
\bibitem{benedeIJ}i P.Bordone, F. Buscemi and C. Benedetti, 
Fluct. Noise Lett. {\bf 11}, 1242003 (2012).
\bibitem{paraoanu14} J. Li, M. P. Silvestri, K. S. Kumar, 
J.-M. Pirkkalainen,A. Veps\"al\"ainen, W. C. Chien, J. Tuorila, M. A. 
Sillanp\"a\"a, P. J. Hakonen, E. V. Thuneberg, and G. S. Paraoanu, 
Nat. Commun. {\bf 4}, 1420 (2013).
\bibitem{bukard} G. Bukard, Phys. Rev. B {\bf 79}, 125317 (2009).
\bibitem{bergli12}H. J. Wold, H. Brox, Y. M. Galperin, and J. Bergli
Phys. Rev. B {\bf 86}, 205404 (2012).
\bibitem{nonmark} C. Benedetti, M. G. A. Paris, and S. Maniscalco, 
Phys. Rev. A {\bf 89}, 012114 (2014).
\bibitem{bylander} J. Bylander, S. Gustavsson, F. Yan, F. Yoshihara, K. Harrabi, G. Fitch,
D. G. Cory, Y. Nakamura, J.-S. Tsai, W D. Oliver, Nat. Phys. {\bf 7}, 565 (2011).
\bibitem{zhang07} J. Zhang, X. Peng, N. Rajendran, and D. Suter, Phys. Rev. A
{\bf 75}, 042314 (2007).
\bibitem{ahmed13}M. A. A. Ahmed, G. A. \'Alvarez, D. Suter, Phys. Rev. A {\bf 87},  042309 (2013).
\bibitem{almog}I. Almog, Y. Sagi, G. Gordon, G. Bensky, G. Kurizki and N. Davidson,  J. Phys. B {\bf 44}, 154006 (2011).
\bibitem{alvarez11} G. A. \'Alvarez, and D. Suter, Phys. Rev. Lett. {\bf 107}, 230501 (2011).
\bibitem{fbm14} M. G. A. Paris, arXiv:1401.4194v2.
\bibitem{benedettiNG} C. Benedetti, F. Buscemi, P. Bordone, M. G. A. Paris, Phys. Rev. A {\bf 89}, 032114 (2014).
\bibitem{lucasz}{\L}. Cywi\'nski, ArXiv:1308.3102v1.
\bibitem{hel76} C. W. Helstrom, Quantum Detection and Estimation Theory
(Academic Press, New York, 1976).
\bibitem{mal93} J. D. Malley, J. Hornstein, Statist. Sci. {\bf 8}, 433 (1993).
\bibitem{brau94} S. Braunstein, C. Caves, Phys. Rev. Lett. {\bf 72}, 3439 (1994).
\bibitem{brody99} D. C. Brody, L. P. Hughston, Proc. Roy. Soc. Lond. A {\bf 454}, 2445
(1998); A {\bf 455}, 1683 (1999).
\bibitem{paris09} M. G. A. Paris, Int. J. Quant. Inf. {\bf 7}, 125 (2009).
\bibitem{esch12} B. M. Escher, L. Davidovich, N. Zagury, R. L. de Matos Filho,
Phys. Rev. Lett. {\bf 109}, 190404 (2012);
B. M. Escher, R. L. de Matos Filho, and L. Davidovich, Braz. J. Phys.
{\bf 41}, 229 (2011).
\bibitem{brida}C. Benedetti, A.P. Shurupov, M.G. A. Paris, G. Brida, and M. Genovese, Phys. Rev. A {\bf 87}, 052136 (2013).
\bibitem{blandino} R. Blandino, M. G. Genoni, J. Etesse, M. Barbieri, M. G. A. Paris, P. Grangier, and R. Tualle-Brouri,
Phys. Rev. Lett {\bf 109}, 180402 (2012).
\bibitem{monras06} A. Monras, Phys. Rev. A {\bf 73}, 033821 (2006).
\bibitem{smerzi07} L. Pezz\'e, A. Smerzi, G. Khoury, J. F. Hodelin, and D. Bouwmeestster,
Phys. Rev. Lett. {\bf 99}, 223602 (2007).
\bibitem{olivares09}S. Olivares and M. G. A. Paris, J. Phys. B {\bf 42}, 055506 (2009).
\bibitem{teklu}B. Teklu, S. Olivares and M. G. A. Paris, J. Phys. B {\bf 42}, 035502 (2009).
\bibitem{kac10} M. Kacprowicz, R. Demkowicz-Dobrzanski, W. Wasilewski, K.
Banaszek, I. A. Walmsley, Nature Phot. {\bf 4}, 357 (2010).
\bibitem{cable10} H. Cable, G. A. Durkin, Phys. Rev. Lett. {\bf 105}, 013603 (2010).
\bibitem{durk10} G. A. Durkin, New J. Phys. {\bf 12} 023010 (2010).
\bibitem{genoni2011} M. G. Genoni, S. Olivares, M. G. A. Paris, Phys. Rev. 
Lett. {\bf 106}, 153603 (2011);
M. G. Genoni, S. Olivares, D. Brivio, S.  Cialdi, D. Cipriani, A.
Santamato, S. Vezzoli, M. G. A. Paris, Phys. Rev. A {\bf 85}, 043817 (2012).
\bibitem{spagnolo2012} N. Spagnolo, C. Vitelli, V. G. Lucivero, 
V. Giovannetti, L. Maccone, and F. Sciarrino, 
Phys. Rev. Lett. {\bf 108}, 233602 (2012). 
\bibitem{mqt3}
T. Rocheleau, T. Ndukum, C. Macklin, J. B. Hertzberg, A. A. Clerk, and
K. C. Schwab, Nature {\bf 463}, 7275 (2010).
\bibitem{mqt4}
A. D. O'Connell, M. Hofheinz, M. Ansmann, R. C. Bialczak, M.
Lenander, E. Lucero, M. Neeley, D. Sank, H. Wang, M. Weides1, J. Wenner,
J. M. Martinis1, A. N. Cleland, Nature {\bf 464}, 697 (2010).
\bibitem{mqt1} {M. Brunelli, S. Olivares, M. G. A. Paris},
Phys. Rev. A {\bf 84}, 032105 (2011).
\bibitem{mqt2} 
{M. Brunelli, S. Olivares, M. Paternostro, M. G. A. Paris}, 
Phys. Rev. A {\bf 86}, 012125 (2012).
\bibitem{dechiara}G. De Chiara and M. Palma, Phys. Rev. Lett. {\bf 91}, 090404 (2003).
\bibitem{mao}J.-W. Mao, J.-X. Chen, W.-H. Huang, B.-Q. Li, W.-K. Ge, Phys. Rev. E {\bf 81}, 031123 (2010).
\bibitem{fiasconaro}A. Fiasconaro and B. Spagnolo, Phys. Rev. E {\bf 80}, 041110 (2009).
\bibitem{benedettiou} C. Benedetti, M. G. A. Paris,  
Int. J. Quantum. Inf. {\bf 12}, 1461004 (2014).
\bibitem{wie24} N. Wiener, Proc. London Math. Soc. {\bf 22}, 454
(1924).
\bibitem{wio13}
 H. S. Wio, {\em Path Integrals for Stochastic Processes} (World
 Scientific, Singapore, 2013). 
\bibitem{leh98} E. L. Lehman, G. Casella, 
{\em Theory of Point Estimation}, (Springer,Berlin, 1998).
\end{thebibliography}
\end{document}